\documentclass[12pt]{article}
\usepackage[dvips]{graphicx}
\usepackage{amsmath}
\usepackage{graphicx}
\usepackage{amsfonts}
\usepackage{amssymb}
\usepackage{epsfig}
\usepackage{subfigure}

\newcommand{\be}{\begin{equation}}
\newcommand{\ee}{\end{equation}}
\newcommand{\bea}{\begin{eqnarray}}
\newcommand{\eea}{\end{eqnarray}}
\newcommand{\ba}{\begin{array}}
\newcommand{\ea}{\end{array}}
\newcommand{\beas}{\begin{eqnarray*}}
\newcommand{\eeas}{\end{eqnarray*}}
\newcommand{\bes}{\begin{equation*}}
\newcommand{\ees}{\end{equation*}}

\def\i2           {\mbox{$\frac{i}{2}$}}

\begin{document}
\title{\bf Dark energy and moduli stabilization of extra dimensions in ${\mathbb
M}^{1+3} \times {\mathbb T}^{2}$ spacetime}

\author{P. Burikham$^{a}$\thanks{Email:piyabut@gmail.com}, A. Chatrabhuti$^{a}$\thanks{Email:auttakit.c@chula.ac.th}, \\
P. Patcharamaneepakorn$^{a}$\thanks{Email:preeda\_patcharaman@hotmail.com}, and K. Pimsamarn $^{b}$\thanks{Email:kpimsa@gmail.com}\\
$^a$ {\small {\em  Theoretical High-Energy Physics and Cosmology Group, Department of Physics,}}\\
{\small {\em Faculty of Science, Chulalongkorn University, Bangkok
10330, Thailand}}\\
$^b$ {\small {\em  Department of Physics, Faculty of Science,
Kasetsart University, Bangkok 10900, Thailand}}}

\maketitle

\begin{abstract}
\noindent Recently, it was found by Greene and Levin that the
Casimir energy of certain combinations of massless and massive
fields in space with extra dimensions play a crucial role in the
accelerated expansion of the late-time universe and therefore it
could serve as a candidate for the dark energy.  It also provides
a mechanism in stabilizing the volume moduli of extra dimensions.
However, the shape moduli of the extra dimensions were never taken
into account in the previous work.  We therefore study the
stabilization mechanism for both volume and shape moduli due to
the Casimir energy in ${\mathbb M}^{1+3} \times {\mathbb T}^{2}$.
The result of our study shows that the previously known local
minimum is a saddle point.  It is unstable to the perturbations in
the direction of the shape moduli.  The new stable local minima
stabilizes all the moduli and drives the accelerating expansion of
the universe.  The cosmological dynamics both in the bulk and the
radion pictures are derived and simulated.  The equations of state
for the Casimir
 energy in a general torus are derived.  Shear viscosity in extra
 dimensions induced by the Casimir density in the late times is
 identified and calculated, it is found to be proportional to the
  Hubble constant. \\


\end{abstract}

\newpage
\section{Introduction}

According to the latest data on Type Ia Supernovae~\cite{sna} and
Cosmic Microwave Background Radiation~(CMBR)~\cite{wmap}, it is
strongly believed that the universe consists of a sort of vacuum
energy, namely dark energy, which contributes the accelerated
expansion in three-dimensional space. Unfortunately, the exact
form of the dark energy has not yet been uncovered until now. The
prominent candidates for dark energy are the cosmological
constant, and models of scalar fields, such as the quintessence
and moduli fields.

In the standard cosmological model where the acceleration of the
universe is taken into account by a positive cosmological constant
term, dark energy contributes largely, more than 70 \% of the
total density of the universe~\cite{wmap}.  This number~(roughly
$10^{-11}$ eV$^{4}$) seems arbitrarily small and the known
mechanisms, such as the popular TeV-scale supersymmetry~(SUSY)
breaking scenario or any top-down high-scale particle physics
mechanisms, fail to produce it.

In recent years, theories with large extra dimensions have
received an explosion of interests as they provide new solution to
the hierarchy problem. Recently, it was found that Casimir energy
of massless and massive fields embedded in higher-dimensional
spacetime could play a crucial role of dark energy with additional
significant properties \cite{Greene,Ponton}. The Casimir energy
not only drives the expansion of universe acceleratedly, but also
stabilizes the volume moduli of extra dimensions. However, the
shape moduli, $\tau_{1},\tau_{2}$, were not included in the work
of Greene and Levin.  In this work we therefore take into account
these moduli in the cosmological dynamics by assuming that the
extra dimensions are ${\mathbb T}^{2}$.  The phenomenological
implications of nontrivial shape moduli were pointed out in
\cite{Dienes,Mafi1,Mafi2}. Shape moduli can have dramatic effects
on the Kaluza-Klein spectrum, for example, they can induce
level-crossings and varying mass gaps. They can also help to
eliminate light KK states.  It should be interesting to
investigate the role of shape moduli in cosmology.

Our work employed the calculation of Casimir energy in the
non-trivial space ${\mathbb M}^{4} \times {\mathbb T}^{2}$. The
Casimir energy is the vacuum energy contributed from the quantum
fluctuation of fields which satisfy certain boundary conditions.
In fact, the Casimir energy in various spaces including a
distorted torus was studied in earlier works
\cite{Ambjorn,Elizalde1,Kirsten,Ponton}. The standard approach for
determining the Casimir energy is the zeta function regularization
\cite{Elizalde2}.

Our result shows that the minimum of potential in the previous
work~\cite{Greene}~($\tau_{1}=0,\tau_{2}=1$) was the unstable
local minimum while the true local minimum locates at specific
points in the moduli space, $\tau_{1}=\pm
1/2,\tau_{2}=\sqrt{3}/2$, confirming the result of
Ref.~\cite{Ponton}.  At this local minimum the potential
stabilizes all moduli and also sources the accelerated expansion
of the four dimensional universe.

This paper is organized as follows.  In Section 2 we review
cosmological dynamics on ${\mathbb M}^{1+n} \times {\mathbb
T}^{p}$ spacetime. In Section 3 we present the mathematical
calculation to determine the Casimir energy of massive and
massless fields in the spacetime with toroidally compactified
extra dimensions.  Then we go on to construct effective potential
contributed by Casimir energy of massive and massless field in
${\mathbb M}^{1+3} \times {\mathbb T}^{2}$ spacetime in section 4.
The numerical evidences of the stability of moduli space are
presented in section 5. In section 6 we present our conclusions.

\section{Cosmological Dynamics in ${\mathbb M}^{1+n} \label{6D}
\times {\mathbb T}^{p}$}

Our study of cosmological dynamics is based upon the application
of Einstein's general relativity on the product space ${\mathbb
M}^{1+n} \times {\mathbb T}^{p}$, between a $(1+n)$-dimensional
spacetime and a $p$-dimensional toroidally-compactified space.  As
a whole, the total number of spatial dimensions is $d = n+p$.  We
assume the cosmological ansatz
\begin{eqnarray}
ds^2 = g_{\mu \nu}(x) dx^{\mu}dx^{\nu} + h_{ij}(x) dy^{i}dy^{j},
\label{g_metric}
\end{eqnarray}
where the metric $h_{i j}$ represent the $p$-dimensional compact
space with $i,j = 1, \cdots, p$ and $g_{\mu \nu}$ for the
$(1+n)$-dimensional noncompact spacetime with $\mu, \nu = 0,
\cdots, n$.  Let's assume also that the metric only depends on the
noncompact coordinates $x^{\mu}$. The compact coordinates are $0
\leq y^i \leq 2\pi$.

In this paper, we focus our effort on the cosmological dynamics of
a 4-dimensional spacetime with two extra dimensions~($n =3$ and $p
=2$).  The metric of two-dimensional torus ${\mathbb T}^2$ takes
the form
\begin{eqnarray}
(h_{i j}) = \frac{b^2}{\tau_2}\left(%
\begin{array}{cc}
  1 & \tau_1 \\
  \tau_1 & |\tau|^2 \\
\end{array}%
\right), \label{h_metric}
\end{eqnarray}
where $\tau = \tau_1+ i \tau_2$ is the complex structure (or shape
moduli) and $b^2$ is the K\"{a}hler structure (or volume moduli).
In cosmology, it is customary to write $g_{\mu\nu} =
a^2(t)\eta_{\mu\nu}$ and $h_{ij} = h_{ij}(t)$.  In the next
sections, we will assume that Casimir energy in compact direction,
$\rho_{(d+1)D}$, plays the roles of the dominant energy content in
the universe.  By using Einstein equations in $(1+5)$-dimensional
spacetime, we obtain the following equations governing the
cosmological dynamics:
\begin{eqnarray}
3H^2_{a} + H_{b}^2 + 6H_{a}H_{b} -
\frac{1}{4\tau^2_{2}}(\dot{\tau}^2_1+\dot{\tau}^2_2) = 8\pi
G\rho_{6D},\label{eom1}\\
\dot{H}_{a}+3H_{a}^2 + 2H_{a}H_{b} = \frac{8\pi
G}{4}\left\{2\rho_{6D} +\left[1 -
\left(\frac{\tau_1}{\tau_2}\right)^2\right]b\partial_{b}\rho_{6D}
- 2\tau_1\partial_{\tau_1}\rho_{6D} +
\frac{2\tau_1^2}{\tau_2}\partial_{\tau_2}\rho_{6D} \right\},\label{eom2}\\
\dot{H}_{b}+2H_{b}^2 + 3H_{a}H_{b} = -\frac{8\pi
G}{4}\left\{-2\rho_{6D} +\left[1 -
\left(\frac{\tau_1}{\tau_2}\right)^2\right]b\partial_{b}\rho_{6D}
- 2\tau_1\partial_{\tau_1}\rho_{6D} +
\frac{2\tau_1^2}{\tau_2}\partial_{\tau_2}\rho_{6D} \right\},\label{eom3}\\
\ddot{\tau}_{1}+\left(3H_a
+2H_b-2\frac{\dot{\tau}_2}{\tau_2}\right)\dot{\tau}_{1} = -16\pi
G\tau_2^2\left\{\frac{b\tau_1}{2\tau_2^2}\partial_{b}\rho_{6D} +
2\partial_{\tau_1}\rho_{6D} -
\frac{\tau_1}{\tau_2}\partial_{\tau_2}\rho_{6D}\right\},\label{eom4}\\
\frac{\ddot{\tau}_{2}}{\tau_2}+
\frac{\dot{\tau}_1^2-\dot{\tau}_2^2}{\tau_2^2}
+3H_a\frac{\dot{b}}{\tau_2} +2H_b\frac{\dot{\tau}_2}{\tau_2} =
8\pi G\left\{\frac{b\tau_1^2}{\tau_2^2}\partial_{b}\rho_{6D} +
2\tau_1\partial_{\tau_1}\rho_{6D} - 2\tau_2\left[1 +
\left(\frac{\tau_1}{\tau_2}\right)^2\right]\partial_{\tau_2}\rho_{6D}\right\}.\label{eom5}
\end{eqnarray}
where $G$ is the 6-D gravitational constant.  We have defined the
Hubble constants $H_a = \dot{a}/a$ and $H_b = \dot{b}/b$, where a
dotted quantity represents the corresponding time derivative and
$\rho_{6D}$ is the casimir energy density in six dimensional
spacetime.

\subsection{Dynamics in the Radion picture}

Equations of motion (\ref{eom1})-(\ref{eom5}) can be obtained by
varying the $d+1$-dimensional Einstein-Hilbert action:
\begin{eqnarray}
S= \int d^{1+n}x d^{p}y \sqrt{-gh}\Big\{
\frac{M^{d-1}_{*}}{16\pi}\mathcal{R}_{(1+d)}-\rho_{(1+d)D}({h^{ij}})\Big\},
\end{eqnarray}
with $n=3$ and $p=2$, where $\rho_{(1+d)D}({h^{ij}})$,
$\mathcal{R}_{(1+d)}$ and $ M_{*}$ are the Casimir energy density,
Ricci scalar and the Planck mass in $(1+d)$-dimensional spacetime
respectively.
 For later purpose, it is useful to perform KK-dimensional reduction of the above action from $(1+d)$ to
$(1+n)$-dimensional spacetime and Weyl rescaling $g_{{\mu \nu_E}}
= \Omega^{\frac{2}{n-1}} g_{\mu \nu}$; $\Omega =
M^{d-1}_{*}V_{p}/m^{n-1}_{pl}$, the action takes the form
\begin{eqnarray}
S = \int d^{1+n}x \sqrt{-g_E} \Big\{
\frac{m^{n-1}_{pl}}{16\pi}\Big[ {\mathcal R}_{E}+g^{\mu
\nu}_E\Big(\frac{1}{1-n}
\nabla_{\mu}\ln\sqrt{h}\nabla_{\nu}\ln\sqrt{h}+\frac{1}{4}\nabla_{\mu}h^{ij}\nabla_{\nu}
h_{ij}\Big)\Big] - U({h^{ij}})\Big\}. \label{reduce_action}
\end{eqnarray}
Note that the subscript $E$ denotes the Einstein frame variables.
Here, $V_p = \int d^p y \sqrt{h}=(2\pi b)^{p}\equiv l^{p}$ is the
(invariant) volume of extra dimensions, $m_{pl}$ and $U(h^{ij}) =
\Omega^{\frac{1+n}{1-n}} V_p \rho_{(1+d)D}(h^{ij}) =
\Omega^{\frac{1+n}{1-n}} \rho_{(1+n)D}(h^{ij})$ are the Planck
mass and the effective potential in $1+n$-dimensional spacetime
respectively. We can also take $\rho_{(1+n)D}(h^{ij})$ to be the
Casimir energy density in $(1+n)$-dimensional spacetime.

Since we are interested in the $n =3$, $p=2$ case, by using the
metric of two-dimensional torus defined in Eqn.~(\ref{h_metric}),
the action in Eqn.~(\ref{reduce_action}) can be written as
\begin{equation}
S = \int d^{4}x \sqrt{-g_E} \Big\{ \frac{m^{2}_{pl}}{16\pi}\Big[
{\mathcal R}_{E}-\frac{1}{2} g^{\mu
\nu}_E(\nabla_{\mu}\psi\nabla_{\nu}\psi+
e^{-2\phi_2}\nabla_{\mu}\phi_1 \nabla_{\nu}
\phi_1+\nabla_{\mu}\phi_2 \nabla_{\nu} \phi_2)\Big] -
U(\psi,\phi_1,\phi_2)\Big\},
\end{equation}
where $\psi \equiv 2\sqrt{2}\ln b$, $\phi_1 \equiv \tau_1$, and
$\phi_2 \equiv \ln \tau_2$.  Such action gives rise to the
following set of equations:
\begin{eqnarray}
6H_E^2-\frac{1}{2}(\dot{\psi}^2+e^{-2\phi_2}\dot{\phi_1}^2+\dot{\phi_2}^2)&=&\frac{16\pi}{m^2_{pl}}U,\\
\ddot{\psi}+3H_E\dot{\psi}&=&-\frac{16\pi}{m^2_{pl}}\frac{\partial U}{\partial \psi},\\
\ddot{\phi_1}+3H_E\dot{\phi_1}-2\dot{\phi_1}\dot{\phi_2}&=&-\frac{16\pi}{m^2_{pl}}e^{2\phi_2}\frac{\partial
U}{\partial \phi_1},\\
\ddot{\phi_2}+3H_E\dot{\phi_2}+e^{-2\phi_2}\dot{\phi_1}^2&=&-\frac{16\pi}{m^2_{pl}}\frac{\partial
U}{\partial \phi_2},
\end{eqnarray}
and
\begin{equation}
4\dot{H}_E +
(\dot{\psi}^2+e^{-2\phi_2}\dot{\phi_1}^2+\dot{\phi_2}^2)  = 0.
\end{equation}
Note that $H_E = (da_E/dt_E)/a_E$ is the Hubble constant in the
Einstein's frame.

\section{Casimir energy in ${\mathbb M}^{1+n} \times
{\mathbb T}^{p}$}

In this section, we will undergo the mathematical formulation to
determine the Casimir energy, $\widehat{E}_{cas}$, associated with
a scalar field of mass $M$ in a ${\mathbb M}^{1+n} \times {\mathbb
T}^{p}$ space.  The fermionic degree of freedom will contribute to
the Casimir energy with the same expression except for an extra
minus sign.  We then focus on the result from our phenomenological
study $(n=3, p=2)$.

\subsection{Casimir-Energy Calculation} \label{rho}

Let $V_{n}=L^{n}$ be the spatial volume of non-compact spacetime,
and $V_{p}=l^{p}$ be the volume of compact space. If we assume $L
\gg l$, the zero-point energy of scalar fields in $ {\mathbb
M}^{1+n} \times {\mathbb T}^{p}$ can be evaluated by
\begin{eqnarray}
\widehat{E}_{cas} &=& \frac{1}{2}(\frac{L}{2 \pi})^{n}
\sum_{n_i,n_j} \int^{+ \infty}_{- \infty} d^{n}k  \sqrt{ \delta^{a
b} k_{a} k_{b}+ h^{ij}n_{i}n_{j}+ M^2},
\end{eqnarray}
where $k_a;$ $a=1,\ldots,n$ is the momentum in each non-compact
spatial direction, $n_{i} \in {\mathbb Z};$ $i=1,\ldots,p$ is the
momentum number in each compact direction.

Using the property of integration in Appendix~\ref{A} and changing
variable of integration as $v= k^{2}/(h^{i j}n_{i}n_{j}+M^2)$, we
can express the Casimir energy as
\begin{eqnarray}
\widehat{E}_{cas} &=& \frac{1}{2}(\frac{L}{2
\pi})^{n}\frac{\pi^{n/2}}{\Gamma(n/2)} \sum_{n_i,n_j}
(h^{ij}n_{i}n_{j}+ M^2)^{\frac{n+1}{2}} \int^{ \infty}_{0} dv
v^{\frac{n-2}{2}}\sqrt{1+v}. \label{eq:cas1}
\end{eqnarray}
We can convert the integral into the Gamma function by using the
formulae in Appendix~\ref{A}; as a consequence, we obtain the
Casimir energy in a simple form
\begin{eqnarray}
\widehat{E}_{cas} &=& \frac{1}{2}(\frac{2 \pi}{L})^{1+2s}
\frac{\Gamma(s)}{\pi^{\frac{1+2s}{2}}\Gamma(-\frac{1}{2})}
\sum_{n_i,n_j} (h^{ij}n_{i}n_{j}+M^2)^{-s};\quad
s=-\frac{d-p+1}{2}.   \label{eq:cas2}
\end{eqnarray}

In our case, the compact space is ${\mathbb T}^{2}$ and $h^{ij}$
is the inverse metric from Eqn.~(\ref{h_metric}); therefore, our
next task is to regularize the infinite summation in the
Eqn.~(\ref{eq:cas2})
\begin{eqnarray}
F(s;\frac{|\tau|^2}{b^2 \tau_2},-\frac{2\tau_1}{b^2
\tau_2},\frac{1}{b^2 \tau_2};M^2) &=& \sum_{n_1,n_2}
(\frac{|\tau|^2}{b^2 \tau_2}n^{2}_{1}-\frac{2\tau_1}{b^2 \tau_2}
n_1 n_2+ \frac{1}{b^2 \tau_2}n^{2}_{2} +M^2)^{-s}, \label{eq:cas3}
\end{eqnarray}
which is known as extended Chowla-Selberg zeta
function~\cite{Elizalde1}.  It is worth noting that $V_p=l^2=(2\pi
b)^2$ in this case.

After a few steps of analytic manipulation by using Poisson
resummation and property of the modified Bessel function, we
obtain
\begin{eqnarray}
F(s;\frac{|\tau|^2}{b^2 \tau_2},-\frac{2\tau_1}{b^2
\tau_2},\frac{1}{b^2 \tau_2};M^2) &=& b^{2s}
\{2\tau^{s}_{2}\zeta_{EH}(s;\tau_{2}b^2M^2)+
 2\sqrt{\pi} \frac{\Gamma(s-\frac{1}{2})}{\Gamma(s)} \tau^{1-s}_{2}
\zeta_{EH}(s-1/2;\frac{b^2M^2}{\tau_{2}}) \label{eq:cas4}\\
\nonumber &&+ \sum^{\infty}_{m,k=1} \frac{8 \pi^{s}}{\Gamma(s)}
\sqrt{\tau_2} k^{s-\frac{1}{2}} \frac{\cos(2\pi\tau_{1}m
k)}{(\sqrt{m^{2}+\frac{b^2M^2}{\tau_2}})^{s-\frac{1}{2}}}
K_{s-\frac{1}{2}}(2\pi\tau_2 k
\sqrt{m^{2}+\frac{b^2M^2}{\tau_2}})\},
\end{eqnarray}
where the Epstein-Hurwitz zeta function $\zeta_{EH}(s;q)$ is
expressed as
\begin{eqnarray}
\zeta_{EH}(s;q) &=& \frac{1}{2} {\sum_{n \in {\mathrm
Z}}}^{\prime}
(n^2+q)^{-s}\nonumber \\
&=& -\frac{q^{-s}}{2}+ \frac{\sqrt{\pi}
\Gamma(s-\frac{1}{2})}{2\Gamma(s)}q^{-s+\frac{1}{2}}
+\sum^{\infty}_{n=1} \frac{2
\pi^{s}q^{-s/2+1/4}}{\Gamma(s)}n^{s-\frac{1}{2}}
K_{s-\frac{1}{2}}(2\pi n \sqrt{q}),
\end{eqnarray}
where the prime at the first sum indicates that the term $n=0$ is
excluded.  A similar expression which manifests the periodicity of
the Casimir energy with respect to $\tau_{1}$ is also given in
Ref.~\cite{ElizaldeO}.

The expression serves as an analytic continuation of the Casimir
energy where $s$ is extended from positive to negative values.
 Inserting Eqn.~(\ref{eq:cas4}) into Eqn.~(\ref{eq:cas2}) and
eliminating the infinite terms due to the pole of $\Gamma(s=-2)$
and $\Gamma(s-1=-3)$ in this case, we conveniently reached the
regularized Casimir energy.  The dropped divergent terms
correspond to the constant total energy and the constant energy
density in the bulk. Both of them do not depend on any parameters
of the torus and therefore can be safely eliminated from the
physically relevant Casimir effects by renormalization. The final
regulated Casimir energy density $\rho(h^{i j})$ in
$(1+3)$-dimensional spacetime can then be expressed as
\begin{eqnarray}
\rho_{4D}(b^2,\tau_1,\tau_2) &=& \frac{\widehat{E}_{cas}}{V_{m}}\nonumber\\
&=& -(4 \pi^2 b^2)^s  \Big\{ 2\tau^s_2 (\tau_2 b^2
M^2)^{-\frac{s}{2}+\frac{1}{4}}\sum^{\infty}_{k=1}k^{s-\frac{1}{2}}K_{s-\frac{1}{2}}(2
\pi k b M\sqrt{\tau_2}) \nonumber \\
&&+ 2 \tau^{1-s}_2 (\frac{
b^2M^2}{\tau_2})^{-\frac{s}{2}+\frac{1}{2}}\sum^{\infty}_{k=1}k^{s-1}K_{s-1}(\frac{2
\pi k b M}{\sqrt{\tau_2}}) \nonumber\\
&&  + 4 \sqrt{\tau_2}
\sum^{\infty}_{k,m=1}k^{s-\frac{1}{2}}\frac{\cos(2 \pi \tau_1 k
m)}{(\sqrt{m^2+\frac{b^2
M^2}{\tau_2}})^{s-\frac{1}{2}}}K_{s-\frac{1}{2}}(2 \pi k \tau_2
\sqrt{m^2+\frac{b^2 M^2}{\tau_2}}) \Big\} . \label{eq:cas5}
\end{eqnarray}
In the case of massless scalar fields $(M = 0)$, the Casimir
energy density becomes
\begin{eqnarray}
\rho_{4D}(b^2,\tau_1,\tau_2) &=& -(4 \pi^2 b^2)^s \Big\{\tau^s_2
\pi^{s-\frac{1}{2}}\Gamma(\frac{1}{2}-s)\zeta(1-2s) +
\tau^{1-s}_2\pi^{s-1}\Gamma(1-s)\zeta(2-2s)\nonumber \\
&\quad& +4 \sqrt{\tau_2}
\sum^{\infty}_{m,k=1}(\frac{k}{m})^{s-\frac{1}{2}}\cos(2\pi
mk\tau_1)K_{s-\frac{1}{2}}(2\pi mk\tau_2) \Big\} . \label{eq:cas6}
\end{eqnarray}
The Casimir density in (1+3+2) dimensions is given by
$\rho_{6D}=\rho_{4D}/(2\pi b)^{2}$.

\begin{figure}[htp]
\centering
\includegraphics[width=0.7\textwidth]{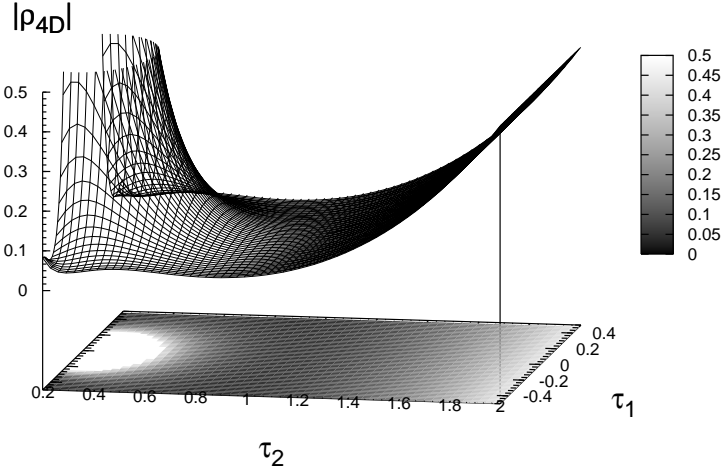}
\caption[$\rho_{6D}$]{The magnitude of the Casimir energy density,
$|\rho_{4D}|$, in four dimension per degree of freedom for $M=5,
b= 0.133$. }\label{c1}
\end{figure}

As it is pointed out in the work of Ponton and Poppitz~\cite{Ponton}.  Since the symmetry $\tau \to -1/\tau, \tau \to \tau +1$ of the
torus is preserved in the Casimir energy expression, it is
sufficient to consider only the fundamental region where $\tau
\geq 1, -1/2\leq \tau_{1}\leq 1/2$ of the shape moduli space.  In
the fundamental region, there are two minima and one saddle point
of the magnitude $|\rho|$ of the Casimir energy density.  The
saddle point locates at $\tau_{1}=0,\tau_{2}=1$ and the two minima
locate at $\tau_{1}=\pm 1/2,\tau_{2}=\sqrt{3}/2$.  This is shown
in Figure~\ref{c1}.

\subsection{Analysis for small $bM$ }

In the limit of $bM \ll 1$, we recalculate the Casimir energy by
performing the binomial expansion with respect to small $bM$
before regularization, and keep only the leading-order terms.  It
can be demonstrated that the process of regularizing each term
after performing binomial expansion is NOT equivalent to the
process of regularizing the whole expression at once if $s=-2$ is
set beforehand.  When we set $s=(1-d)/2 = -2$, the binomial
expansion of Eqn.~(\ref{eq:cas3}) gives only three terms with
orders of $(bM)^{0}, (bM)^{2}$, and $(bM)^{4}$, whereas the
regularization of the full expression before setting $s=-2$ as in
Eqn.~(\ref{eq:cas4}) ,which gives Eqn.~(\ref{eq:cas5}) as a
result, generically leads to an infinite series of $bM$, even
after setting $s=-2$ in the final expression.

Without setting $s=-2$ before regularization, the precise
dependence of the coefficients of the $bM$-binomial expansion to
the moduli parameters $\tau_{1}, \tau_{2}$ will be determined.
The small $bM$ expansion is obtained subsequently.

We begin by replacing $h^{i j}$ with the form of the inverse
metric of ${\mathbb T}^{2}$ in Eqn.~(\ref{eq:cas2}) and using
Mellin transform~(see Appendix \ref{A})
\begin{eqnarray}
\widehat{E}_{cas} &=& \frac{1}{2}(\frac{2 \pi}{L})^{1+2s}
\frac{\Gamma(s)}{\pi^{\frac{1+2s}{2}}\Gamma(-\frac{1}{2})}
\sum_{n_1,n_2 \in {\mathbb Z}} \int^{\infty}_{0} dt~ t^{s-1}
e^{-\{ \frac{1}{\tau_2 b^2} (|\tau|^2 n^{2}_{1} -2 \tau_1
n_{1}n_{2}
+n^{2}_{2}) +M^2 \} t} \nonumber \\
&=& \frac{1}{2}(\frac{2 \pi}{L})^{1+2s}
\frac{\Gamma(s)}{\pi^{\frac{1+2s}{2}}\Gamma(-\frac{1}{2})}
\sum_{n_1,n_2 \in {\mathbb Z}} \int^{\infty}_{0} dv~ v^{s-1}
e^{-\{ \frac{1}{\tau_2} (|\tau|^2 n^{2}_{1} -2 \tau_1 n_{1}n_{2}
+n^{2}_{2}) +(b M)^2 \} v} \nonumber \\
&=& \frac{1}{2}(\frac{2 \pi}{L})^{1+2s}
\frac{b^{2s}}{\pi^{\frac{1+2s}{2}}\Gamma(-\frac{1}{2})}
\sum^{\infty}_{j=0} \frac{(-1)^m}{m!} (b M)^{2j} \Gamma(s+j)
\sum_{n_1,n_2 \in {\mathbb Z}} \left(\frac{|\tau|^2}{\tau_2}
n^{2}_{1} -2 \frac{\tau_1}{\tau_2} n_{1}n_{2} +
\frac{1}{\tau_2}n^{2}_{2}\right)^{-(s+j)}, \nonumber \\
\label{eq:sbm1}
\end{eqnarray}
where the second line is obtained by changing the dummy variable
$v=t/b^{2}$, and the final line is obtained by expanding the
Taylor series for $e^{-(bM)^{2}}$. We can determine the double
summation in Eqn.~(\ref{eq:sbm1}) by using the result in
Eqn.~(\ref{eq:cas3}),(\ref{eq:cas4}); as a consequence, the
Casimir energy density in five spatial dimensions takes the form,
\begin{eqnarray}
\rho_{6D}(b^2,\tau_1,\tau_2)= -(4 \pi^2 b^2)^{s-1}
\sum^{\infty}_{j=0}\frac{(-1)^j}{j!}(b M)^{2j} \Big\{ 4\pi^{j}
\sqrt{\tau_2} \sum^{\infty}_{m,k=1}
(\frac{k}{m})^{s+j-\frac{1}{2}} \cos(2 \pi m
k \tau_1) K_{s+j-\frac{1}{2}}(2 \pi m k \tau_2)  \nonumber \\
 +\pi^{s+2j-\frac{1}{2}} \tau^{s+j}_2
\Gamma(\frac{1}{2}-s-j)\zeta(1-2s-2j)+\pi^{s+2j-1}\tau^{1-s-j}_2
\Gamma(1-s-j)\zeta(2-2s-2j) \Big\}. \label{eq:sbm2} \nonumber
\end{eqnarray}

In the limit $bM \ll 1$ for $s=-2$, the Casimir energy density
then becomes
\begin{eqnarray}
\rho_{6D}(b^2,\tau_{1},\tau_{2}) \simeq - \frac{1}{(4 \pi^{2}
b^{2})^{3}} \Big\{ C_{1} - C_{2}(b M)^{2} + C_{3}(b M)^{4} \Big\}
\label{eq:sbm3}
\end{eqnarray}
where \bea C_{1} & \equiv &
\pi^{-\frac{5}{2}}\tau^{-2}_{2}\Gamma\left(\frac{5}{2}\right)\zeta(5)+\pi^{-3}\tau^{3}_2
\Gamma(3)\zeta(6)+4
\sqrt{\tau_{2}}\sum^{\infty}_{m,k=1}\left(\frac{m}{k}\right)^{\frac{5}{2}}
\cos(2\pi mk \tau_{1}) K_{-5/2}(2\pi mk \tau_{2}), \nonumber \\
C_{2} & \equiv & \pi^{-\frac{1}{2}} \tau^{-1}_{2} \Gamma
\left(\frac{3}{2}\right)\zeta(3)+\pi^{-1}\tau^{2}_2
\Gamma(2)\zeta(4)+4 \pi \sqrt{\tau_{2}}\sum^{\infty}_{m,k=1}\left(
\frac{m}{k}\right)^{\frac{3}{2}} \cos(2\pi mk \tau_{1})
K_{-3/2}(2\pi mk \tau_{2}), \nonumber \\
C_{3} & \equiv &
\frac{\pi}{2}\tau_{2}\Gamma(1)\zeta(2)+2\pi^{2}\sqrt{\tau_{2}}\sum^{\infty}_{m,k=1}
\left(\frac{m}{k}\right)^{\frac{1}{2}}\cos(2\pi
mk\tau_{1})K_{-1/2}(2\pi mk \tau_{2}). \nonumber \\
\label{eq:sbm4} \eea

In the next section, the total Casimir density for small $bM$ and
the full expression will be numerically compared.  The true
minimum of the potential, induced from the Casimir energy density
located at a point $(\tau_{1},\tau_{2})=(\pm 1/2,\sqrt{3}/2)$,
appears only when the full expression is evaluated.

\section{Particle spectrum and effective potential for moduli
fields}


It is demonstrated in Ref.~\cite{Ponton} and Ref.~\cite{Greene}
that a careful mixing of massless and massive, bosonic and
fermionic degrees of freedom of the bulk fields can lead to a
Casimir energy density with local minimum with respect to the
scale factor, $b$, of the compact extra dimensions.  In the torus
case with the shape moduli $\tau_{1}, \tau_{2}$, it can be shown
that the true minimum of the mixed Casimir energy density~(and
thus the potential) locates at $\tau_{1}=\pm 1/2,
\tau_{2}=\sqrt{3}/2$, in contrast to the case of undistorted torus
considered in the previous work where the shape moduli are set to
$\tau_{1}=0,\tau_{2}=1$.

The simplest model of the bulk fields in our ${\mathbb M}^{1+3}
\times {\mathbb T}^{2}$ space consists of a massless boson, a
massless fermion, a massive fermion with mass $M$, and a massive
boson with mass $\lambda M$.  It was found that for the range
$0.40<\lambda<0.42$ and $M=5$, the mixed Casimir density has local
minimum with respect to the scale factor $b$, and the moduli
$\tau_{1},\tau_{2}$.  Since the mass of the boson is different
from the mass of the fermion, this is the scenario where SUSY is
broken in the bulk if it exists at higher scales. There is no
particular reason for why the ratio of the masses of the massive
boson and fermion took the specific value in this range. If it has
anything to do with SUSY breaking, it is desirable that we are
able to establish a SUSY breaking mechanism where this specific
ratio of the masses $\lambda$ could be explained or distinctively
selected. From phenomenological point of view, it is desirable
that these massless and small-mass bulk fields are sterile
neutrinos for they can explain the smallness of neutrino masses in
four dimensions. For further details, see
Ref.~\cite{addm},\cite{cmy}.

\begin{figure}[htp]
\centering
\includegraphics[width=0.7\textwidth]{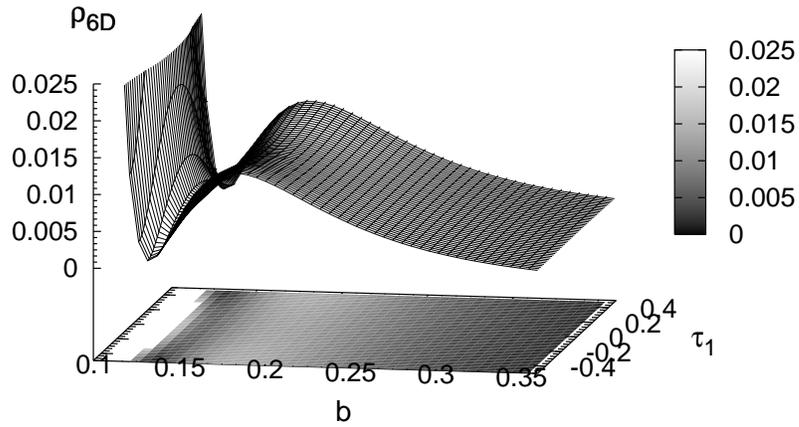}
\caption[$\rho_{6D}$]{The total Casimir energy density in six
dimension for mixture of massless and massive fields for $M=5,
\lambda = 0.408$, and $\tau =\sqrt{\tau_{1}^{2}+\tau_{2}^2}$ is
fixed to $1$. }\label{c2}
\end{figure}

An important issue in mixing bosonic and fermionic degrees of
freedom to obtain the total Casimir energy density with a local
minimum is the positivity of the energy density.  Generally, the
value of the total Casimir density at $\tau_{1}=\pm 1/2,
\tau_{2}=\sqrt{3}/2$ is lower than the value at the saddle point
$\tau_{1}=0, \tau_{2}=1$, for all range of $\lambda$.  However,
for certain ranges of $\lambda$ ({\it e.g.} $\lambda \lesssim
0.407$), the density becomes negative around the true minimum and
therefore violates the positive energy condition.  A negative
value of the density will not stabilize the dynamics and the size
of the torus.  We therefore choose the value $\lambda =0.408$ for
our simulation of the cosmological dynamics.  Figure~\ref{c2}
shows the total Casimir energy density for the spectrum of
massless and massive particles mentioned above.

\begin{figure}[htp]
\centering
\includegraphics[width=0.7\textwidth]{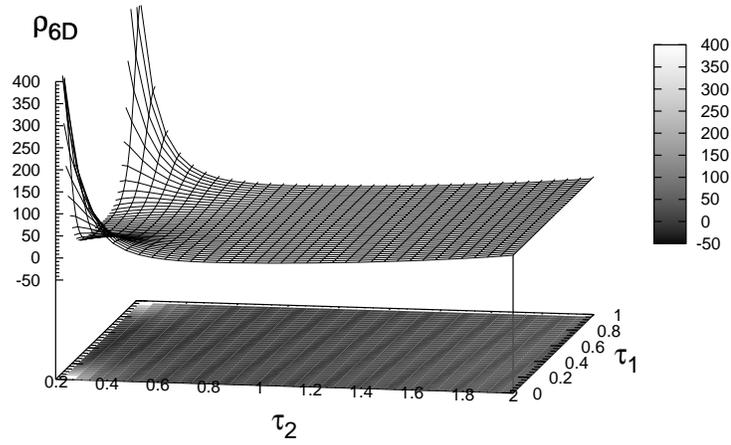} \hfill
\includegraphics[width=0.7\textwidth]{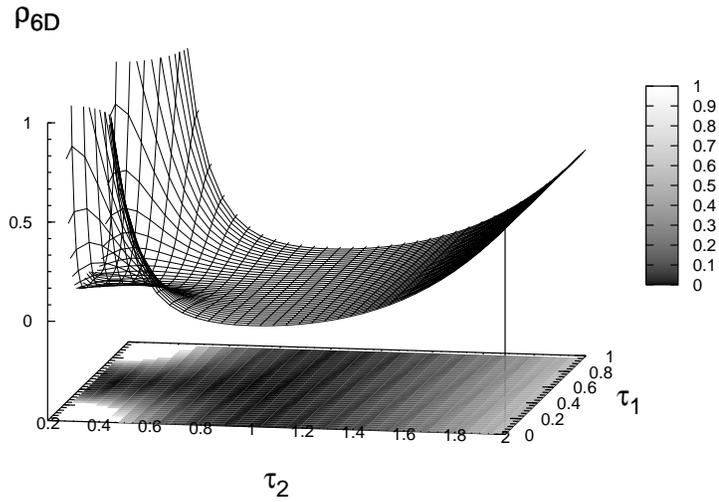}\\
\caption[$\rho_{6D}^{approx},\rho_{6D}^{full}$]{The Casimir energy
density in six dimension from small $bM$ approximation in the
upper figure in comparison to the full expression in the lower
figure.  Both are evaluated at their corresponding $b_{min}$.
}\label{c23}
\end{figure}

The plot of the total Casimir density in (1+3+2)-dimensional
spacetime using the full expression, Eqn.~(\ref{eq:cas5}), in
comparison to the plot from the small $bM$ approximation,
Eqn.~(\ref{eq:sbm3}), is given in Figure~\ref{c23}.  The true
minimum at $\tau_{1}=\pm 1/2, \tau_{2}=\sqrt{3}/2$ only exist in
the full expression case.  This can be understood considering
$b_{min}M \approx 0.67$ and is somewhat close to $1$, resulting in
a bad approximation of the expression due to higher powers of $bM$
being neglected.  It is therefore required that we use the full
expression of the total Casimir energy density in the simulation
of the cosmological dynamics.

\section{Evidence of stability of the moduli space and cosmological dynamics}

By numerically solving the field equations in section~\ref{6D},
the stabilization of the torus and the accelerated expansion of
large 4-dimensional spacetime can be demonstrated to occur at the
true minimum of the Casimir energy density in the moduli space.
The point $\tau_{1}=0,\tau_{2}=1$ is a saddle point and it is an
unstable equilibrium of the dynamics.

\begin{figure}
     \centering
     \subfigure[$a(t)$]{
          \includegraphics[width=.45\textwidth]{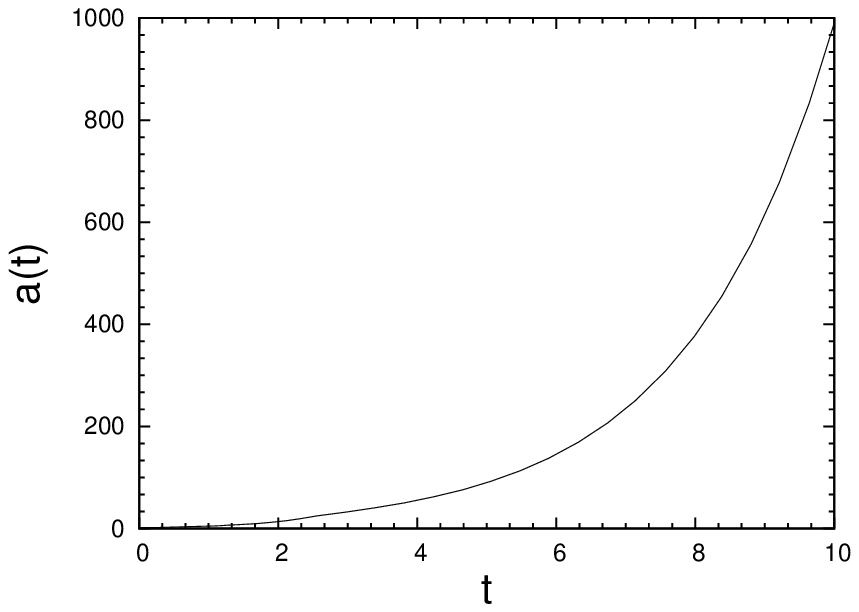}}
     \subfigure[$b(t)$]{
          \includegraphics[width=.45\textwidth]{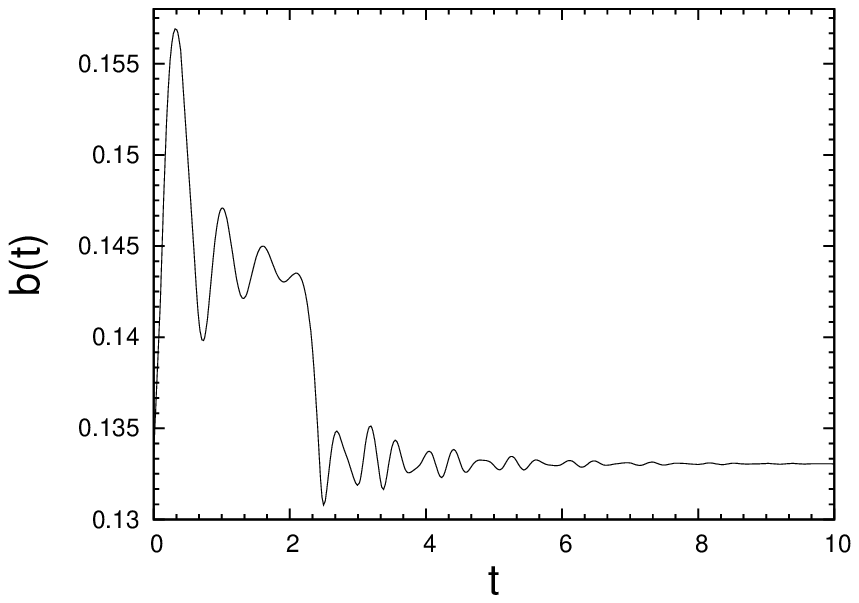}}\\
     \vspace{.3in}
     \subfigure[$H_{a}$]{
           \includegraphics[width=.45\textwidth]
                {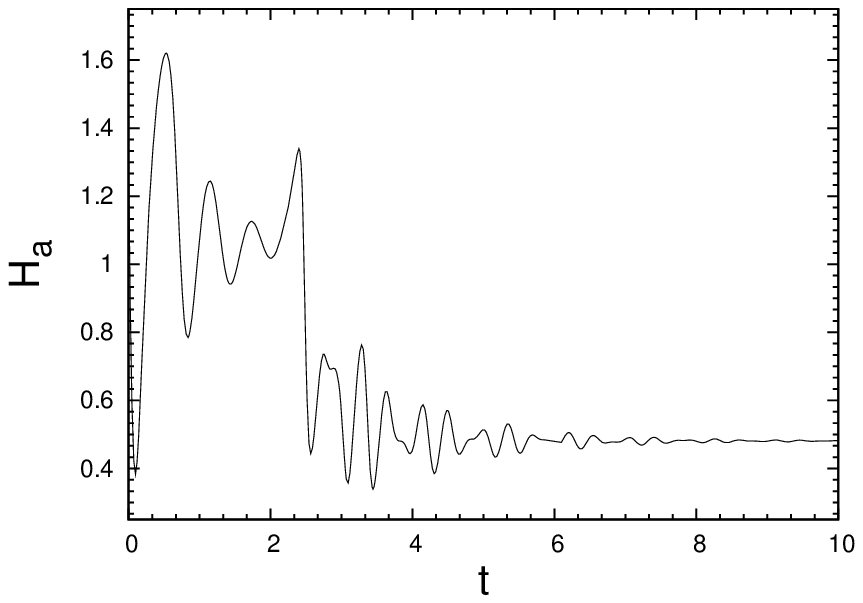}}
     \subfigure[$H_{b}$]{
          \includegraphics[width=.45\textwidth]{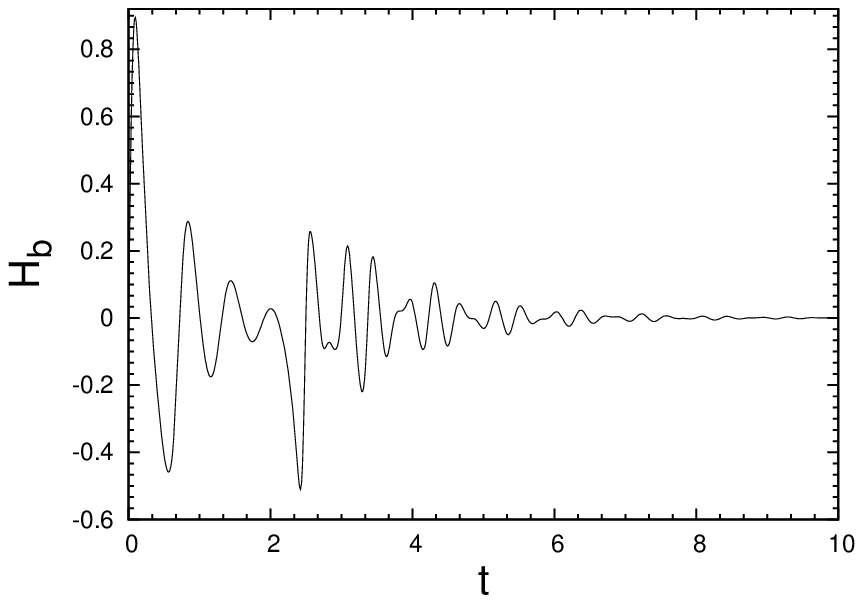}}
     \vspace{.3in}
     \subfigure[$\tau_{1}$]{
           \includegraphics[width=.45\textwidth]
                {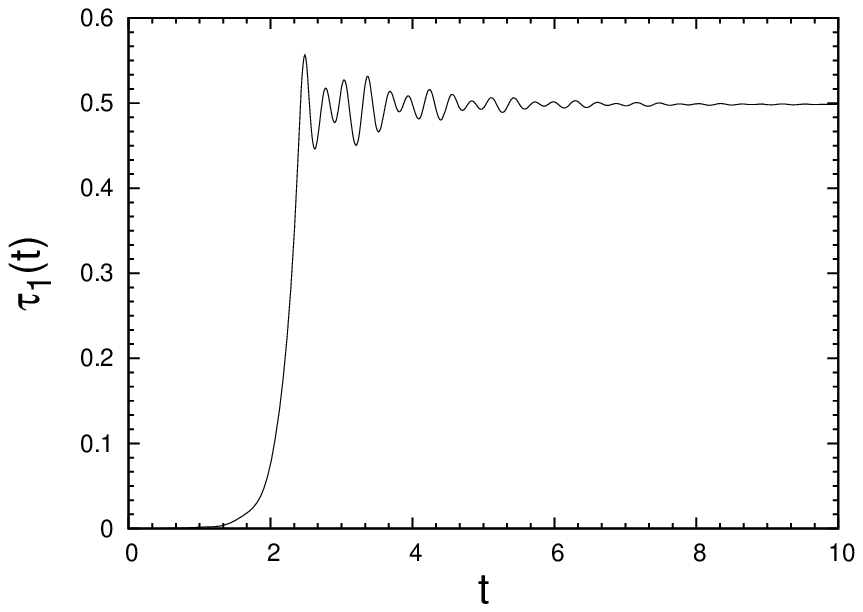}}
     \subfigure[$\tau_{2}$]{
          \includegraphics[width=.45\textwidth]{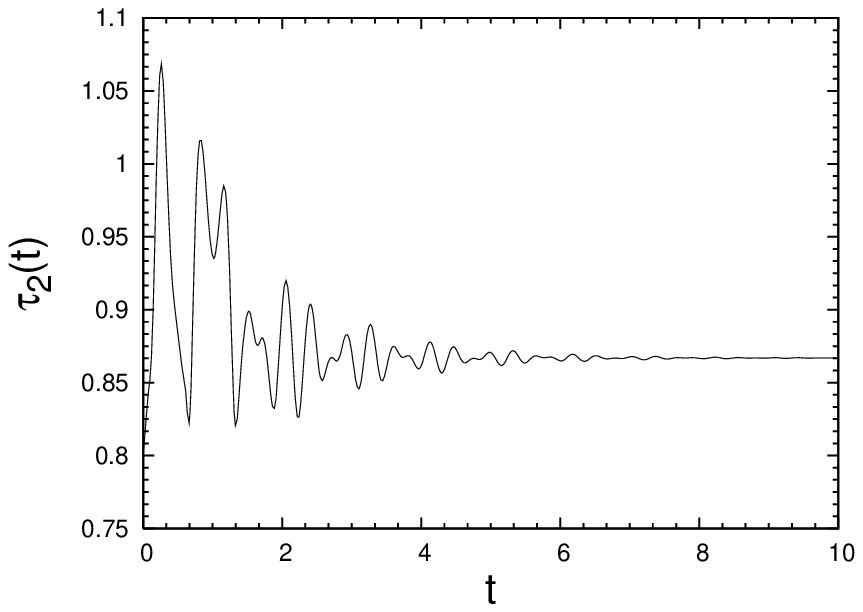}}
      \caption{Cosmological dynamics when the universe is initially
tossed very close to the saddle point $\tau_{1}=0,\tau_{2}=1$, it
rolls along the trail $\tau =1$ to the true minimum at
$\tau_{1}=\pm1/2, \tau_{2}=\sqrt{3}/2$.}
     \label{r1}
\end{figure}

\begin{figure}
     \centering
     \subfigure[$\tau$ trail]{
          \includegraphics[width=.45\textwidth]{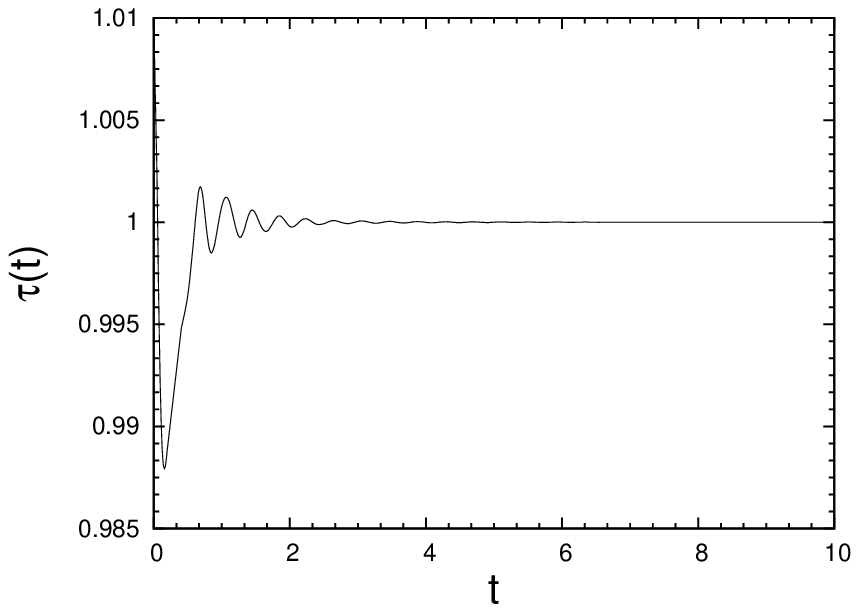}}
     \subfigure[$w_{b}$]{
          \includegraphics[width=.45\textwidth]{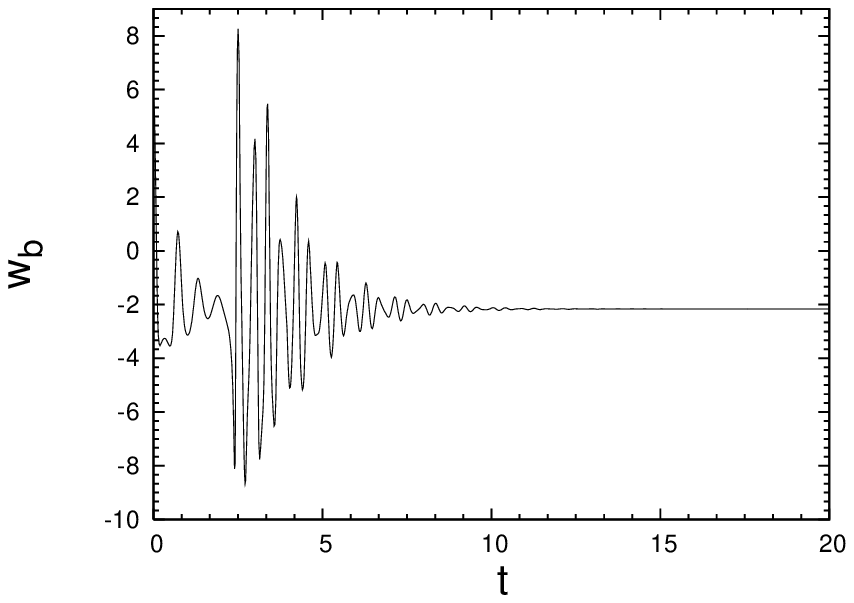}}\\
     \subfigure[$w_{4}$]{
          \includegraphics[width=.45\textwidth]{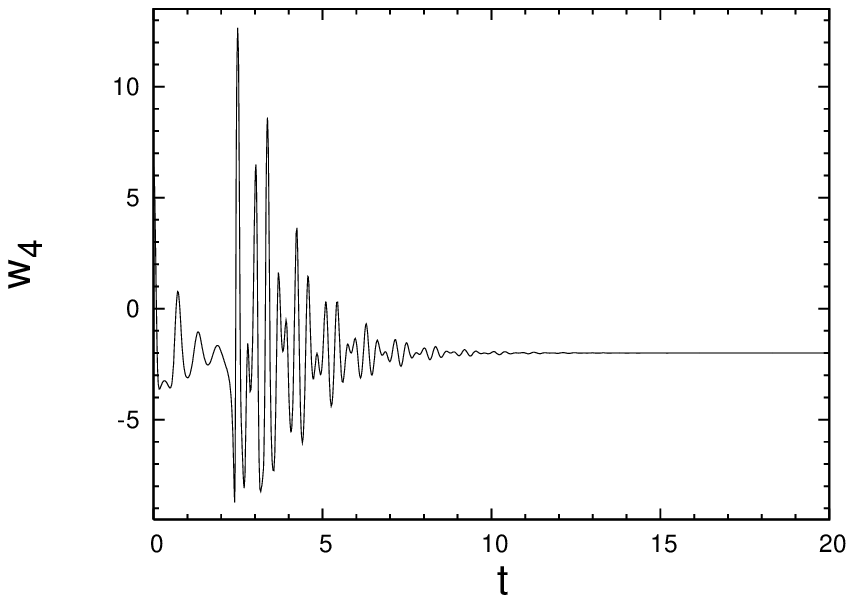}}
     \subfigure[$w_{5}$]{
          \includegraphics[width=.45\textwidth]{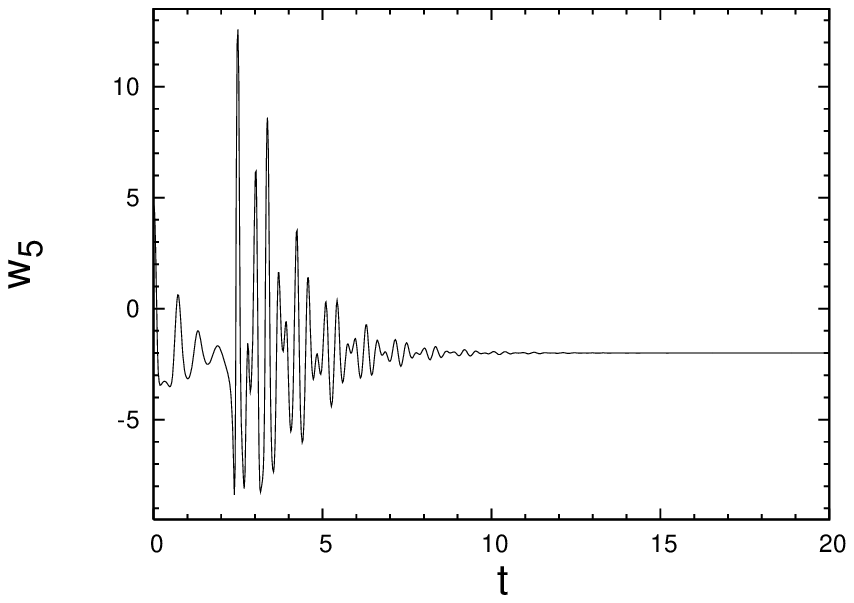}}\\
     \subfigure[rolling in shape moduli plane]{
          \includegraphics[width=.45\textwidth,height=2.5in]{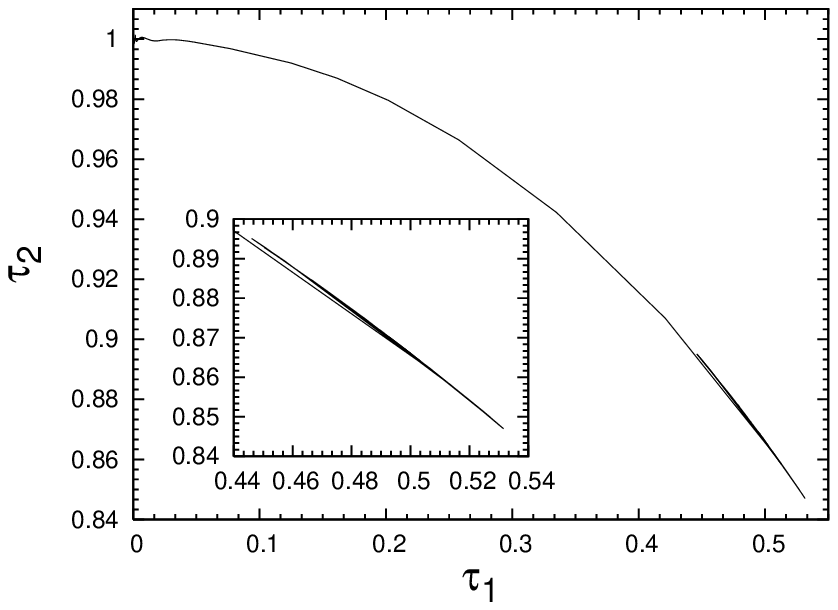}}
     \subfigure[rolling in the landscape]{
          \includegraphics[width=.48\textwidth,height=2.8in]{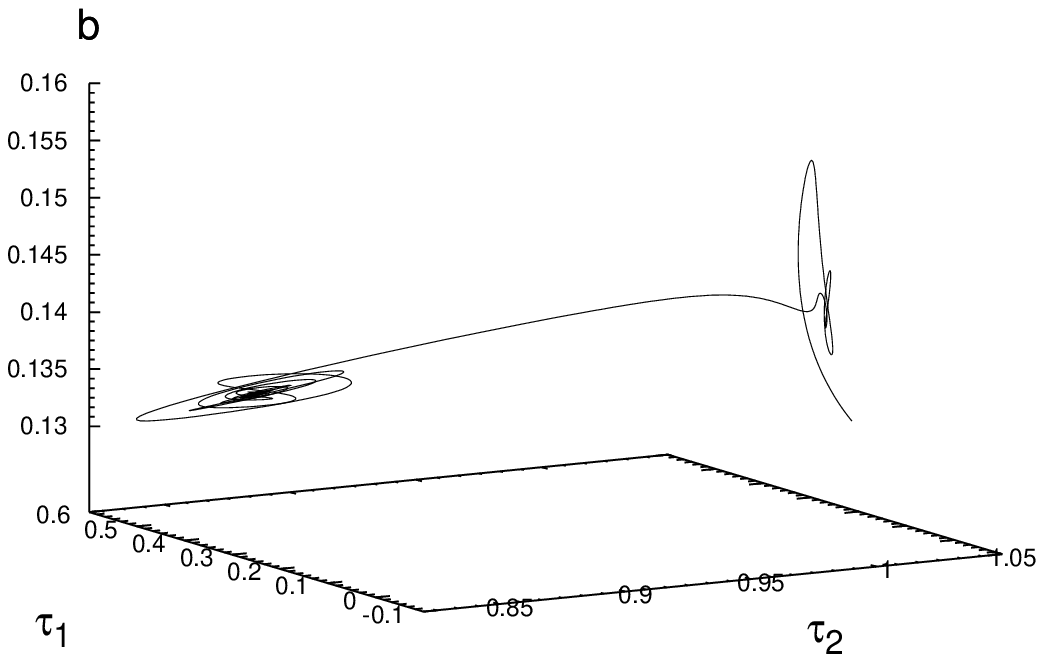}}\\
     \caption{Rolling dynamics from saddle point to the true minimum.}
     \label{r12}
\end{figure}

\begin{figure}
\centering
\includegraphics[width=.7\textwidth,height=3.5in]{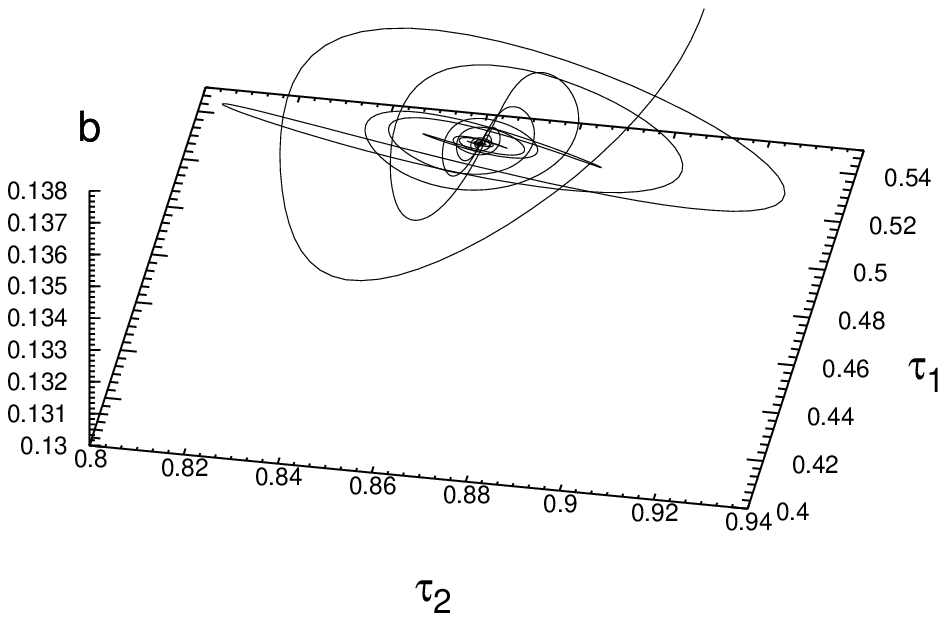}
\caption[I]{Rolling dynamics from other initial condition I.}
\label{r2}
\end{figure}

\begin{figure}
\centering
\includegraphics[width=.7\textwidth,height=3.5in]{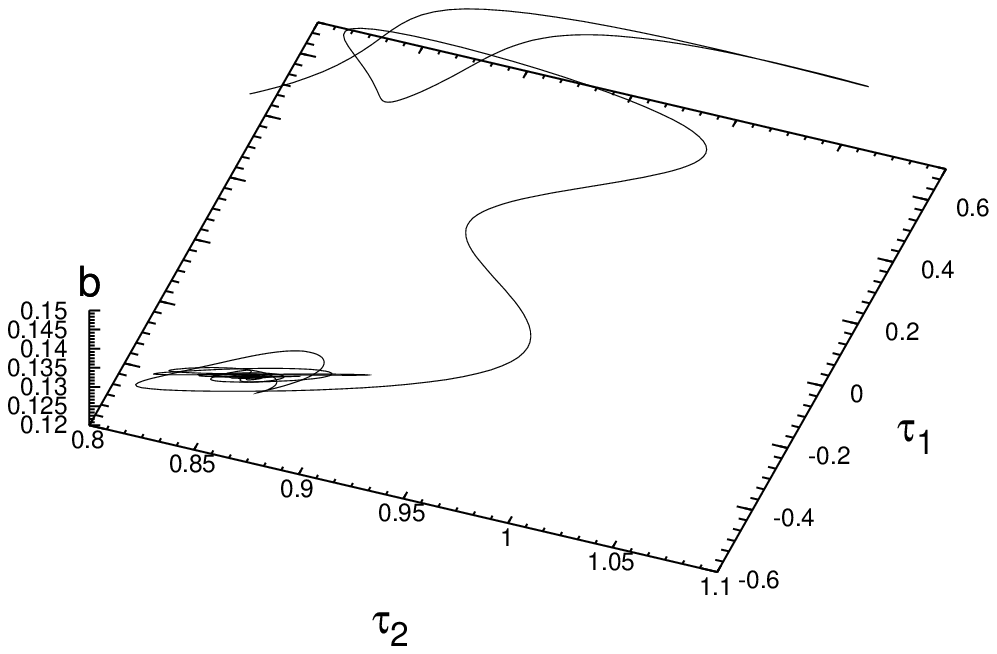}
\caption[II]{Rolling dynamics from other initial condition II.}
\label{r3}
\end{figure}

The rolling of the universe to the true minimum of the Casimir
density is illustrated in Figure~\ref{r1}-\ref{r3}. When the
cosmological dynamics is initiated even within a small vicinity of
the saddle point, $\tau_{1}=0,\tau_{2}=1$, of the Casimir energy
density, it will roll down to the true minimum at $\tau_{1}=\pm
1/2, \tau_{2}=\sqrt{3}/2$ even with minimal amount of
perturbations. This is shown in Figure~\ref{r1}, \ref{r12}.
Observe that it tends to roll along the trail $\tau =1$ in the
moduli space.

When the tossing initial conditions are at a distant away from the
saddle point and the true minimum, certain sets of the initial
conditions still result in the stabilization of the torus moduli,
$\tau_{1},\tau_{2}$, and the scale factor, $b$, of the extra
dimension as is shown in Figure~\ref{r2}, \ref{r3}.  Naturally, as
long as the Casimir energy density at the stabilized value is
positive, the acceleration of the scale factor, $a$, of the
4-dimensional spacetime is guaranteed.  The positive Casimir
density serves as the positive cosmological constant.

A natural consequence of the Casimir energy that is independent of
the scale factor, $a(t)$, of the large dimension is the fact that
it leads to $w_{a}=-1$ for the pressure $p_{a}=w_{a}\rho$.  For
the pressure in the compact extra dimensions, we can start by
considering $p_{b}=-\partial (\rho V_{b})/\partial
V_{b}=w_{b}\rho$, $w_{b}$ of our Casimir energy density is then
given by
\begin{eqnarray}
w_{b}&=& -1-\frac{b}{2\rho}\frac{\partial \rho}{\partial b}
\end{eqnarray}
where $\rho$ is the total Casimir energy density.  Due to the
dynamics of shape moduli~(or Casimir ``viscosity" in the compact
space, see Appendix \ref{B}), the value of $w_{b}$ at the
stabilized radius at the true minimum is fractionally smaller than
$-2$~(around $-2.16$) as is shown in Figure~\ref{r12}.

A more appropriate definition of physical pressures in the
distorted torus is
\begin{eqnarray}
p^{*}_{K}&\equiv & T^{K}_{K},
\end{eqnarray}
where $K=4,5$.  This definition gives the following expressions
for $w_{K}=p^{*}_{K}/\rho$,
\begin{eqnarray}
w_{4}&=& -1 +
\frac{b}{2\rho}\partial_{b}\rho(\frac{\tau_{1}^{2}-\tau_{2}^{2}}{\tau_{2}^{2}})
+\frac{2\tau_{1}}{\rho}\partial_{\tau_{1}}\rho +
\frac{1}{\rho}(\frac{\tau_{2}^{2}-\tau_{1}^{2}}{\tau_{2}})\partial_{\tau_{2}}\rho \\
w_{5}&=& -1 +
\frac{b}{2\rho}\partial_{b}\rho(\frac{\tau_{1}^{2}-\tau_{2}^{2}}{\tau_{2}^{2}})
-
\frac{1}{\rho}(\frac{\tau^{2}}{\tau_{2}})\partial_{\tau_{2}}\rho.
\end{eqnarray}
By directly solving the equations of motion in six dimensions at
the stabilized point where $\dot{H}_a =
\dot{H}_b=H_b=\dot{\tau}_1=\dot{\tau}_2=0$, it can be shown that
$w_{4,5}= -2$, as is confirmed numerically in Figure~\ref{r12}. It
is interesting to note that the value of $w_{4,5}$ becomes $-2$ at
both the saddle point and the true minimum where the dynamics is
stabilized.

The difference of the two definitions of pressure originates from
the {\it shear viscosity} induced by the Casimir energy in the
off-diagonal components of the stress tensor.  From the equations
of motion of the 6-D universe with viscosities, Eqn.~(\ref{veom4})
in Appendix~\ref{B}, shear viscosity at the stabilized point
$\eta_{b}^{stab}$ can be identified to be
\begin{eqnarray}
\eta_{b}^{stab}&=& \frac{3H_{a,stab}}{16\pi G} \\
               &=& \frac{\rho_{6D,min}}{2H_{a,stab}}
\end{eqnarray}
where $H_{a,stab}$ is the Hubble constant of the expanding four
dimensions at the stabilized point of the compactified space. Note
that we can evaluate Eqn.~(\ref{eom1}),~(\ref{veom2}) and
(\ref{veom3}) at the stabilized point and use the definition of
$\eta_b^{stab}$ to analytically confirm the numerical results in
which $w_{4,5} = -2$ at the stabilized point.

We should mention here that the time scale, $t_{s}$, of the
simulated figures is given by
\begin{eqnarray}
t_{s} &=& \frac{\sqrt{23}}{2 \pi}\frac{m_{pl}}{b_{min}}
~b_{s}^{3},
\end{eqnarray}
where $b_{s}$ is the scale of $b$, and $b_{min}\simeq 0.1328
b_{s}$ as a result of numerical simulation.  If we require that
the stabilization time $\simeq 10 t_{s}$ is less than the age of
the universe, $10^{10}$ years, this will put constraint on the
size $b_{min}$ of the extra dimensions ${\mathbb T}^{2}$,
\begin{eqnarray}
b_{min}&\lesssim& 0.7 \mu\mbox{m}.
\end{eqnarray}
This is about few hundred times stronger than the constraints from
table-top experiments~\cite{exp}.

It is interesting that in this kind of cosmological model, the
constancy of the 4-dimensional gravitational constant,
$G_{4}=G/4\pi^{2}b^{2}=1/m_{pl}^{2}$, up to the early times of the
universe will give a very strong constraint on the size of the
compactified extra dimensions.  Any future observations of the
universe from very early epoch could possibly put constraints on
the inconstancy of the gravitational constant.  Such constraints
will put very strong limits on the size of compact extra
dimensions in this kind of model where oscillatory behaviour is
significant in the early times.

Another important aspect of this model is the relationship between
the effective cosmological constant in 4-dimensional spacetime,
$\Lambda_{4} = 8\pi G_{4} \rho_{4D,min}$, and the size of extra
dimension, $b_{min}$,
\begin{eqnarray}
\Lambda_{4} & = & 8\pi G_{4} \rho_{4D,min}  \\
        & = & 3 H_{E,stab}^{2}.
\end{eqnarray}
This leads to the typical value of $b_{min}\approx 2.4 ~\mu$m for
$\rho_{vac}\approx 10^{-11}$ eV$^{4}$.  The value of the effective
size of extra dimensions, $2 \pi b_{min}\approx 15 \mu$m, yields
the quantum gravity scale in the bulk, $M_{*}\approx 12$ TeV.

\section{Conclusions and discussion}

The stabilization of compact extra dimensions and the acceleration
of the other 4-dimensional part of the spacetime can be
simultaneously described by the dynamics of the Einstein field
equations in the bulk spacetime. The acceleration of the
4-dimensional ``universe" occurs naturally once the scale of the
compact dimensions is stabilized and the density of the Casimir
energy in the bulk becomes a (positive) constant at that
stabilized value. As a result, the apparent positive
``cosmological constant" that we seem to observe in the four
dimensional visible universe is effectively induced.  This is
demonstrated beautifully in the work by Greene and
Levin~\cite{Greene} when the Casimir density of the undistorted
torus satisfies $w_{a}=-1,w_{b}=-2$ condition.

Shape moduli of the torus can be added to the model.  The true
minimum of the Casimir energy density of the torus with shape
moduli is demonstrated to be located at $\tau_{1}=\pm 1/2,
\tau_{2}=\sqrt{3}/2$.  The cosmological dynamic shows that a
minimally small perturbation to the saddle point rolls the
universe down to the true minimum.  Other initial conditions also
suggest that the universe tends to roll around $\tau=1$ contour to
reach the true minimum. Note that it is also possible to stabilize
the moduli at the saddle point $\tau_1 = 0$, $\tau_2 = 1$ but the
initial conditions of the shape moduli fields must be fine-tuned
so that $\tau_{1}=0, \dot{\tau_{1}}=0$.  Some extra-mechanisms
such as Brandenberger-Vafa mechanism in string gas cosmology
\cite{Brandenberger} is needed for this purpose.  However, as it
was pointed out in \cite{auttakit}, the stabilized point $\tau_1 =
\pm 1/2$ and $\tau_2 = \sqrt{3}/2$ is also the fixed point of
T-duality and the the enhance symmetry point hence
Brandenberger-Vafa mechanism could also set the initial value of
the moduli precisely to be at the stabilized point.

The shear viscosity in the extra dimension is determined to be
proportional to the Hubble constant at the stabilized point,
$\eta_{b}=3H_{a,stab}/16\pi G$.  Through the Einstein field
equations, this Hubble constant of the 4-D universe is determined
by the value of the Casimir energy density at the stabilized
point.  The effective four dimensional cosmological constant is
also given by $8\pi G \rho_{6D,min}$.

In this kind of model, there is a relationship between the size of
the compact dimensions and the observed four dimensional
cosmological constant.  This remarkable connection is induced by
the nature of Casimir energy density which depends on the size of
the compact dimension, resulting in $\Lambda_{4} \sim
b_{min}^{-6}$.

It is equally important to note that the constancy of the 4-D
gravitational constant up to very early time of the universe will
provide strong constraint on the size of extra dimension in this
particular cosmological model which expresses oscillatory
behaviour at the early times.

\label{conclusions-sec}

\section*{Acknowledgments}
\indent We would like to thank Khamphee Karwan and Aphisit
Ungkitchanukit for valuable discussions.  A.C. is supported in
part by the Thailand Research Fund (TRF) and Commission on Higher
Education (CHE) under grant MRG5180225.

\appendix

\section{Useful Formulae} \label{A}

{\bf Phase Space Integration}
\begin{eqnarray}
\int d^{d}k f(k)&=& \frac{2 \pi^{d/2}}{\Gamma(d/2)}\int
k^{d-1}f(k)dk.
\end{eqnarray}

{\bf Poisson Resummation}
\begin{eqnarray}
\sum_{n \in \mathbb{Z}}f(n) = \tilde{f}(k) &=& \sqrt{2\pi}\sum_{m
\in \mathbb{Z}} \tilde{f}(2\pi m),
\end{eqnarray}
where
\begin{eqnarray}
\tilde{f}(k) &=& \frac{1}{\sqrt{2\pi}} \int^{+\infty}_{-\infty}
f(x)e^{-ikx} dx.
\end{eqnarray}

If $f(x) = e^{-a(x+c)^2}$, then $ \tilde{f}(k) =
\frac{1}{\sqrt{2\pi}}\sqrt{\frac{\pi}{a}}e^{-\frac{k^2}{4a}+ikc}$.\\

{\bf Integral representation of Gamma function}
\begin{eqnarray}
\Gamma(x) = \int^{+ \infty}_{0} e^{-t} t^{x-1} dt.
\end{eqnarray}
{\bf The integral representation of the modified Bessel function
of the second kind}
\begin{eqnarray}
K_{\nu}(z) = \frac{1}{2} \Big(\frac{z}{2}\Big)^{\nu}
\int^{+\infty}_{0} t^{-\nu-1} e^{-t-\frac{z^2}{4t}} ~dt,
\end{eqnarray}
where $|arg(z)|< \frac{\pi}{2}$,$\mathrm{Re}(z^2)>0$.

{\bf Mellin transform}
\begin{eqnarray}
z^{-s} &=& \frac{1}{\Gamma(s)} \int^{\infty}_{0}dt ~e^{-z
t}t^{s-1}; \quad \mathrm{Re}(z)>0,\quad \mathrm{Re}(s)>0.
\end{eqnarray}

\section{Energy momentum tensor of viscous fluid} \label{B}

Let $U^{A} = (1,0,0,0,0,0)$ be the 6-velocity of the cosmic fluid
in comoving coordinates. In terms of the projection tensor
$h_{AB}=g_{AB}+U_{A}U_{B}$, the general energy momentum tensor of
fluid with bulk viscosity $\zeta$ and shear viscosity $\eta$ is
given by:
\begin{equation}
T_{AB} = \rho U_A U_B + (p-\zeta\theta)h_{AB}-2\eta\sigma_{AB}.
\end{equation}
Here $\theta\equiv\nabla_{A}U^{A}$ is the scalar expansion and
$\sigma_{AB} =
h^C_Ah^D_B\nabla_{(C}U_{D)}-\frac{1}{5}h_{AB}\theta$ is the shear
tensor.  By using metric defined in Eqn.~(\ref{g_metric}) and
(\ref{h_metric}), we can show that
\begin{eqnarray}
T^0_0 &=& -\rho, \\
T^1_1 &=& T^2_2 = T^3_3 = p_a\\
T^4_4 &=&
(p_b-\zeta_b\theta)-2\eta_b[\frac{3}{5}(H_b-H_a)-\frac{\dot{\tau}_2}{2\tau_2}-\frac{\tau_1\dot{\tau}_1}{2\tau_2^2}]\\
T^5_5 &=&
(p_b-\zeta_b\theta)-2\eta_b[\frac{3}{5}(H_b-H_a)+\frac{\dot{\tau}_2}{2\tau_2}+\frac{\tau_1\dot{\tau}_1}{2\tau_2^2}]\\
T^4_5 &=&
2\eta_b[\frac{\tau_1\dot{\tau}_2}{\tau_2}+(\tau_1^2-\tau_2^2)\frac{\dot{\tau}_1}{2\tau_2^2}]\\
T^5_4 &=& -\eta_b\frac{\dot{\tau}_1}{\tau_2^2}
\end{eqnarray}
Here we assume there is no viscosity in noncompact large
dimensions ($\zeta_a = \eta_a = 0$).  Einstein's equations,
Eqn.~(\ref{eom2})-(\ref{eom5}), can be written in terms of bulk
and shear viscosity as
\begin{eqnarray}
\dot{H}_{a}+3H_{a}^2 + 2H_{a}H_{b} &=& \frac{8\pi
G}{4}\left\{\rho_{6D} + p_a - 2(p_b-\zeta_b\theta)+\frac{12}{5}\eta_b(H_b-H_a)\right\},\label{veom2} \nonumber \\ \\
\dot{H}_{b}+2H_{b}^2 + 3H_{a}H_{b} &=& \frac{8\pi
G}{4}\left\{\rho_{6D} -3p_a + 2(p_b-\zeta_b\theta) - \frac{12}{5}\eta_b(H_b-H_a) \right\},\label{veom3} \nonumber \\ \\
\ddot{\tau}_{1}+\left(3H_a
+2H_b-2\frac{\dot{\tau}_2}{\tau_2}\right)\dot{\tau}_{1} &=& 16\pi
G\left\{\eta_b\dot{\tau}_1\right\},\label{veom4}\\
\frac{\ddot{\tau}_{2}}{\tau_2}+
\frac{\dot{\tau}_1^2-\dot{\tau}_2^2}{\tau_2^2}
+3H_a\frac{\dot{b}}{\tau_2} +2H_b\frac{\dot{\tau}_2}{\tau_2} &=&
48\pi G\left\{\eta_b\frac{\dot{\tau}_2}{\tau_2}
\right\}.\label{veom5}
\end{eqnarray}
The conservation of energy is
\begin{eqnarray}
\dot{\rho}_{6D} + 3H_a(\rho_{6D}+p_a)
&+&2H_b(\rho_{6D}+p_b)\nonumber\\
+(\frac{12}{5}\eta_b-6\zeta_b)H_aH_b&-&(\frac{12}{5}\eta_b+4\zeta_b)H_b^2-\eta_b(\frac{\dot{\tau}_1^2}{\tau_2^2}+\frac{\dot{\tau}_2^2}{\tau_2^2})=0.
\end{eqnarray}

\begin{thebibliography}{11}
\bibitem{sna} W.L. Freedman {\it et al.}, Astrophys.\ J.\ Suppl.\ {\bf
553} (2001) 47
[arXiv:astro-ph/0012376].
\bibitem{wmap} G. Hinshaw {\it et al.}, Astrophys.\ J.\ Suppl.\ {\bf
170} (2007) 288 [arXiv:astro-ph/0603451]; L. Page {\it et al.},
Astrophys.\ J.\ Suppl.\ {\bf 170} (2007) 335
[arXiv:astro-ph/0603450]; N. Jarosik {\it et al.}, Astrophys.\ J.\
Suppl.\ {\bf 170} (2007) 263 [arXiv:astro-ph/0603452]; D.N.
Spergel {\it et al.}, Astrophys.\ J.\ Suppl.\ {\bf 170} (2007) 377
[arXiv:astro-ph/0603449].
\bibitem{Greene} B.R. Greene and J. Levin, JHEP {\bf 0711} (2007) 096
[arXiv:0707.1062].
\bibitem{Ponton} E. Ponton and E. Poppitz, JHEP {\bf 0106} (2001) 019
[arXiv:hep-ph/0105021].
\bibitem{Dienes} K.R. Dienes, Phys.\ Rev.\
Lett.\ {\bf 88} (2002) 011601 [arXiv:hep-ph/0108115].
\bibitem{Mafi1} K.R. Dienes and A. Mafi, Phys.\ Rev.\ Lett.\ {\bf 88} (2002)
011602 [arXiv:hep-th/0111264].
\bibitem{Mafi2} K.R. Dienes and A. Mafi, Phys.\ Rev.\ Lett.\ {\bf 89} (2002)
171602 [arXiv:hep-ph/0207009].
\bibitem{Ambjorn} J. Ambjorn and S. Wolfram, Annals of Physics {\bf 147}, 1-32 (1983).
\bibitem{Elizalde1} E. Elizalde, J.\ Phys.\ {\bf A27} (1994) 3775-3786
[arXiv:hep-th/9402155].
\bibitem{Kirsten} K. Kirsten and E.
Elizalde, Phys.\ Lett.\ {\bf B365} (1996) 72
[arXiv:hep-th/9508086].
\bibitem{Elizalde2} E. Elizalde, S.D.
Odintsov, A. Romeo, A.A. Bytsenko, and S. Zerbini, \emph{Zeta
regularization techniques with applications}. World Sci.,
Singapore, 1994.
\bibitem{ElizaldeO} E. Elizalde, R. Kantowski, and
S.D. Odintsov, Phys.\ Rev.\ {\bf D54}, (1996) 6372-6380
[arXiv:hep-th/9601101].
\bibitem{addm} N. Arkani-Hamed, S. Dimopoulos, G. Dvali, and J.
March-Russell, Phys.\ Rev.\ {\bf D65}, 024032 (2002)
[arXiv:hep-ph/9811448].
\bibitem{cmy} D.O. Caldwell, R.N. Mohapatra, and S.J. Yellin,
Phys.\ Rev.\ {\bf D64}, 073001 (2001)
[arXiv:hep-ph/0102279].
\bibitem{exp} C.D. Hoyle {\it et al.}, Phys.\ Rev.\ Lett.\ {\bf
86}, (2001) 1418 [arXiv:hep-ph/0011014]; J. Chiaverini {\it et
al.}, Phys.\ Rev.\ Lett.\ {\bf 90}, 151101 (2003)
[arXiv:hep-ph/0209325]; J.C. Long {\it et al.}, Nature {\bf 421},
922 (2003).
\bibitem{Brandenberger} R.H. Brandenberger, {\it Moduli stabilization in string gas cosmology}, arXiv:hep-th/0509159.
\bibitem{auttakit} A. Chatrabhuti, Int.\ J.\ Mod.\ Phys.\ {\bf A22} (2007) 165-180,
[arXiv:hep-th/0602031].

\end{thebibliography}
\end{document}